\documentclass[conference]{IEEEtran}

\ifCLASSOPTIONcompsoc
  \usepackage[nocompress]{cite}
\else
  \usepackage{cite}
\fi
\usepackage{blindtext, graphicx}
\usepackage{forest}
\usepackage{listings}
\usepackage{algorithm}
\usepackage[noend]{algpseudocode}
\usepackage{caption}
\usepackage{pdfcomment}
\usepackage{tcolorbox}
\usepackage{balance}

\algdef{SE}[SUBALG]{Indent}{EndIndent}{}{\algorithmicend\ }%
\algtext*{Indent}
\algtext*{EndIndent}

\usepackage{hyperref}

\usepackage{array}

\usepackage[T1]{fontenc}

\begin{document}

\title{Plug and Play Bench: Simplifying Big Data Benchmarking Using Containers }

\author{\IEEEauthorblockN{Sheriffo Ceesay and Adam Barker}
\IEEEauthorblockA{School of Computer Science\\University of St Andrews, UK\\
Email: \{sc306, adam.barker\}@st-andrews.ac.uk
}
\and
\IEEEauthorblockN{Blesson Varghese}
\IEEEauthorblockA{School of Electronics, Electrical Engg. and Computer Science\\Queen's University Belfast, UK\\
Email : varghese@qub.ac.uk}}

\maketitle

\begin{abstract}
The recent boom of big data, coupled with the challenges of its processing and storage gave rise to the development of distributed data processing and storage paradigms like MapReduce, Spark, and NoSQL databases. With the advent of cloud computing, processing and storing such massive datasets on clusters of machines is now feasible with ease. However, there are limited tools and approaches, which users can rely on to gauge and comprehend the performance of their big data applications deployed locally on clusters, or in the cloud. Researchers have started exploring this area by providing benchmarking suites suitable for big data applications. However, many of these tools are fragmented, complex to deploy and manage, and do not provide transparency with respect to the monetary cost of benchmarking an application.

In this paper, we present Plug And Play Bench (PAPB\footnote{Details of the implementation, README and source code can be obtained from https://github.com/sneceesay77/papb}): an infrastructure aware abstraction built to integrate and simplify the deployment of big data benchmarking tools on clusters of machines. PAPB automates the tedious process of installing, configuring and executing common big data benchmark workloads by containerising the tools and settings based on the underlying cluster deployment framework. Our proof of concept implementation utilises HiBench as the benchmark suite, HDP as the cluster deployment framework and Azure as the cloud platform. The paper further illustrates the inclusion of cost metrics based on the underlying Microsoft Azure cloud platform.

\end{abstract}

\IEEEpeerreviewmaketitle

\section{Introduction}
\begin{figure}[h!]
    \centering
    \includegraphics[scale=0.47]{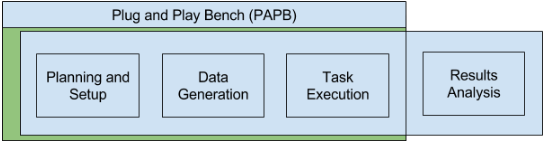}
    \caption{ Big Data Benchmarking Pipeline }
    \label{fig:bmpipeline}
\end{figure}

The public cloud computing market has grown dramatically both in providers and adoption in recent years. Major players like Amazon, Microsoft, Google, and RackSpace provide various software and bare-bone hardware for companies and individuals to rent on a pay as you go basis. Cloud computing can save costs in terms of acquiring the hardware and administering that infrastructure. The ubiquity of this phenomena presents several challenges. Users are often inundated with various service options available from a particular cloud service provider. This problem scales when a user wants to compare different cloud providers. Each provider has different criteria for renting their services, with different billing systems across each of these services. As a result, end users who want to rent a single or cluster of machines to run big data applications are not only challenged with the problems of which cloud service provider to adopt, but also the best configurations of virtual machines for cost-effective usage, or optimal performance. Hence researchers have now started developing tools and methodologies to compare these infrastructures across different cloud service providers. Figure \ref{fig:bmpipeline} illustrates the general pipeline used by current big data benchmarking approaches. The planning and setup stage configures the benchmark suite for execution. This configuration may involve setting up environment variables, setting up data size to generate etc. The second stage involves synthetic data generation and the third stage involves execution of the workload on the cluster. 

This paper focuses on the automation of the first three components, and is primarily motivated by the challenges of big data benchmarking associated with the planning and setup stages, which are discussed below: 

\begin{itemize}\label{challenges}
\item Fragmented Cluster Benchmark Suites, Tools and Approaches: Currently there are various intertwined benchmark suites aimed at benchmarking big data systems. Some of these efforts reinvent the wheel by implementing what has already been implemented in other tools (HiBench and BigDataBench) and sometimes in a less efficient manner. We understand that this may be normal, however, it will be much more appropriate to have a wholly integrated implementation that can be easily deployed and used. 
\item Complex to deploy: We have deployed HiBench, BigDataBench in an 8 node cluster on Microsoft Azure and some technical and domain knowledge is needed in order to configure and run the benchmarks. Source files have to be downloaded, compiled and each benchmark has some specific configurations to set. This includes, data to generate, setting Hadoop and spark home directories, setting executor memory configuration for spark's optimal performance. 
\item Understanding the cost of benchmarking: Finally in a cloud deployment setting, users will most likely be interested to know how costly will it be to run their application in the cloud. Current big data benchmarking tools don't include cost metrics, their main metrics are execution time and throughput. 
\end{itemize}

For a faster and wider adoption of these cluster benchmark tools and suites by the big data and cloud computing community, it is important for their deployment and execution to be simplified for use by both the technical and non-technical users. The main contribution of this work utilises docker containers \cite{merkel2014docker} to implement a proof of concept infrastructure aware benchmarking tool, which simplifies the deployment and usage of current the current tools and suite; and the inclusion of cloud monetary cost metrics. It is, however, important to note that this implementation uses HiBench and HDP but the same idea can be used to integrate other benchmarking tools.


The remainder of this paper is organised as follows. In Section \ref{relatedwork}, we present related works done in comparing and contrasting big data benchmarking approaches. In section \ref{stacsbench}, we present the details of PAPB, finally section \ref{conclusion} covers future work and conclusion.

\section{Related Work}\label{relatedwork}
Figure \ref{fig:bmpipeline} shows the common approach adopted big data benchmarking suites and therefore this section is going to consider how each step of this pipeline is approached by the suites that are evaluated. We consider this rather long and detailed related work important, because it may help readers understand the current state of the art of comprehensive benchmark tool, their pros and cons. 
\begin{figure}[h!]
    \centering
    \includegraphics[scale=0.5]{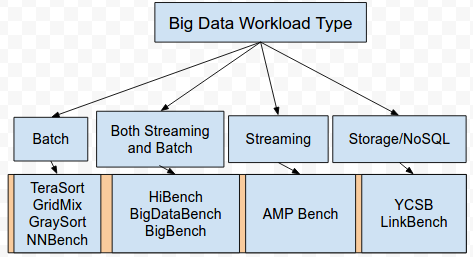}
    \caption{ Common Big Data Benchmark Groupings or Categories }
    \label{fig:tree}
\end{figure}
Figure \ref{fig:tree} shows a high-level view of how current benchmarking tools and suites are categorised. Big data systems are mainly divided into two main categories and they are processing/analytic and storage. The first category of benchmarks focuses on data processing and analytic using batches or stream processing frameworks like Hadoop, Spark or Flink. Some tools and suites focus solely on one of these systems while others aim to provide a comprehensive suite to handle both types of systems. For example, TeraSort which is one of the most common benchmarking tools distributed with Hadoop sorts 1TB of data (or any other amount of data you want) as quickly as possible. It combines testing the HDFS and MapReduce layers of a Hadoop cluster \cite{tera-sort}. Some approaches to Benchmarking like HiBench, BigDataBench and BigBench take a more comprehensive approach by including both batch and stream processing workloads. AMP Benchmarks \cite{ampbench} were created by the authors of Spark and it includes spark, hive, and shark-related benchmarks. The second major big data systems are the NoSQL databases which include document-based stores like MongoDB, wide column stores like HBase, key-value stores like Redis and graph models like voltdb. The most common comprehensive benchmark suite for NoSQL database is YCSB \cite{Cooper:2010:BCS:1807128.1807152} which was developed by Yahoo focusing on HBase, Yahoo! PNUTS and Cassandra but it has now been extended by the open-source community to include benchmarks for other NoSQL databases. Linkbench was developed by Facebook also aims to benchmark workloads involving graph-based data. Table \ref{tab:1} summarises the key properties of each of the main cluster benchmarking tools. Although there are other types of big data benchmarking tools and suites, we selected these four because they are more comprehensive in their approach. For example, \cite{huang2010hibench} and \cite{wang2014bigdatabench} covers lot of workloads in different domains in the analytics space (e.g. Micro bench, Machine Learning, SQL, Web Search and Graph workloads) while \cite{Cooper:2010:BCS:1807128.1807152} covers lot of databases in the NoSQL space (e.g. MongoDB, HBase, Cassandra, Yahoo! PNUTS etc.). In the remainder of this section, we will focus on the approaches used by four comprehensive benchmarking suites that we have investigated. We will first provide a general overview of their approaches, then the metrics used, workload coverage, deployment strategies, diversity of data used and data generation method, big data software stacks covered and finally a critique of their suite.


\subsection{HiBench}
HiBench \cite{huang2010hibench} developed by Intel is a representative, comprehensive and open source big data benchmark suite for Hadoop, Spark, and Streaming Workloads. It consists of workloads for different big data systems and a data generator that generates synthetic data of configurable sizes for those workloads. The workloads are comparably well abstracted and easily configurable to generate and run benchmarks. HiBench is out of the box compatible with all major distribution of Hadoop/spark/streaming frameworks e.g. Apache, Cloudera, and Hortonworks distributions. 

\subsubsection{Big data software stack used}
HiBench utilises the Hadoop Ecosystem which includes software stacks and frameworks like MapReduce, Storm, Flink, Nutch, Hive. Form most Hadoop workloads, it also implements the equivalent of its Spark implementation.   

\subsubsection{Metrics Used}
The metrics used in HiBench are the following: execution time of a workload, throughput and system resource utilisation. At the end of running a workload, all these metrics are appended to an output file for analysis. HiBench also provides a web-based output that displays system resource usage for each executed workload.

\subsubsection{Workloads Used and Their Diversity}
As of writing this paper, the current version of HiBench is 6.0 with 17 workloads categorised into six major areas. The micro-bench category includes both Hadoop and Spark implementations. Some of the included workloads in this category are : word count, a CPU intensive program which counts the number of words in a file; sort, a memory intensive program that sorts an input text file; TeraSort, which is a standard benchmark created by Jim Gray \cite{o2008terabyte}; Sleep, which is a MapReduce program that test the framework's scheduler and DFSIO which stress test the Hadoop Distributed File System by executing lot of reads and writes. The second category of workloads in \cite{huang2010hibench} are classified as Machine Learning workloads. These include Bayesian classification, KMeans clustering, and logistic regression. The third category focuses on SQL Analytics using Hive with workloads on scans, joins, and aggregations. The next category focuses on Graph workloads and includes NWeight which is an iterative graph-parallel algorithm implemented by Spark GraphX and pregel \cite{malewicz2010pregel}. The algorithm computes associations between two vertices that are n-hop away. The Websearch category which performs analytics on web-based data includes the popular PageRank algorithm \cite{page1999pagerank} and Nutch Indexing as their workload. Finally, the streaming category contains various streaming workloads. Example the Identity workloads read input data from Kafka and writes the results back to Kafka while Stateful Wordcount cumulatively counts words received from Kafka every few seconds.

\subsubsection{Data Generation}
For micro-benchmarks, HiBench uses a configurable data generator which mostly used Hadoop's RandomTextWriter to write binary text directly into HDFS. For Web Search and Bayesian Classification, HiBench uses Wikipedia page-to-page link database and Wikipedia dump file respectively. For KMeans they have developed a random data generator using statistic distribution for data generation. Hi-Bench mostly generates unstructured datasets targeting the Hadoop file system.

\subsubsection{Deployment Strategy}
Source code can be downloaded, compiled and deployed in a Unix operating system. Major components like Hadoop and Spark are configured using configuration files. 

\subsubsection{Critique}
HiBench only covers Hadoop and Spark base systems. NoSQL databases which are normally used to store Big Data for easy of processing like HBase, MongoDB, ArangoDB etc. are not included. 

\subsubsection{Pros}
HiBench is fairly simple to install, configure and use. It also outputs the metrics in a very organised manner. 

\subsection{BigDataBench}
BigDataBench \cite{wang2014bigdatabench} is the outcome of a joint research effort between the University of Chinese Academy of Science, Dropbox, Yahoo!, Tencent, Huawei, and Baidu. It consists of a data generator and a benchmark suite. Their main goal is to provide benchmark capability for a diverse set of big data application and dataset types. In their more recent release of the benchmark suite, they used a concept of \cite{gao2015identifying} "big data dwarf" to mitigate the challenge of representing all possible big data computing workloads. Their idea is to construct a benchmark suite using the minimum set of units of computation to represent the diversity of big data analytics workloads.

\subsubsection{Big Data Software Stacks Used}
All major big data software stacks were included in their benchmark suite. This includes Hadoop and Spark Ecosystem. MySQL was also used to benchmark read, write and scans compared to HBase.

\subsubsection{Metrics Used}
Metrics used in BigDataBench are divided into two major categories: user-perceivable-metrics and architectural metrics. The former category includes request per second, operations per second and data processed per second ( RPS, OPS, and DPS  ) which are used to measure the throughput of online services, cloud OLTP workloads, and offline services respectively. Latency or time taken is also used as part of user-perceivable metrics. The latter category is used to compare workload from different categories. 

\subsubsection{Workloads Used and Their Diversity}
As of writing this survey paper, the benchmark suite has 33 big data workload grouped into five big data and application domains and they are: Search Engine, Social Network, E-Commerce, Multimedia Analytics and Bioinformatics. Each of these categories uses a mix of algorithms and big data software stacks. Example, the PageRank workload is a micro-benchmark in the search engine domain which was tested using Hadoop, Spark, Flink, and MPI. Similarly, KMeans which they classified in the social media domain uses Hadoop, Spark, Flink and Spark Streaming.

\subsubsection{Data Generator}
A big data generator \cite{ming2013bdgs} suite was developed to generate various kinds of data in a scalable manner preserving the 4V properties. It is designed for a wide class of application domains which includes search engines, social network, e-commerce, multimedia, and bioinformatics. The data generator also covers the generation of structured, semi-structured and unstructured data types. 

\subsubsection{Deployment Strategy}
Deployment involves downloading and compiling the source can be downloaded and compiled with ease. Hadoop and other big data environment are configurable in configuration files. The benchmarks can be executed on a cluster of Hadoop. and spark clusters. 

\subsubsection{Critique}
Data Generator not efficient, data is locally generated and then copied over the network to 
HDFS. This could be improved by directly writing data to HDFS as the data is generated.
Grouping of the workloads is not very logical e.g. KMeans is classified in the Social Network domain but can also be used in other domains like E-commerce. The metrics that are supposed to be returned are not written or displayed in a console after a workload has been executed. This has to be implemented by the user. 

\subsubsection{Pros}
BigDataBench covers of a lot of application domains, data-types i.e. structured, semi-structured and unstructured. Like HiBench the project is open-source making it possible to extend and test their system.

\subsection{BigBench}
BigBench \cite{wang2014bigdatabench} is an end-to-end big data benchmark proposal the underlying design of which follows a business model of product retailer. BigBench is not an open-source project and therefore not available for deployment and testing by others. Their workloads were executed on Teradata Aster DBMS which is based on nCluster; a shared-nothing parallel database optimised for data warehouse and analytic workloads

\subsubsection{Software Stack Used}
BigBench implements their concept on Teradata Aster Database. SQL-MR queries were used to run the benchmarks. 

\subsubsection{Metrics Used}
The only metrics evidently used in their work is execution time.

\subsubsection{Workloads Used and Their Diversity}
Thirty queries were used in their benchmark which focuses on a business model of product retailer. From a technical perspective, the workloads or queries used spans the following categories: data sources which include structured, semi-structured and unstructured, the second technical focus is query processing types which constitute procedural and declarative data processing paradigm and the third one is analytic techniques which include statistical analysis, data mining and simple reporting. As for diversity, their work presents only a case for product retailer base workloads.

\subsubsection{Data Generator}
BigBench used the Parallel Data Generation Framework (PDGF) which was designed to address structured data generation. They have extended this framework to cater for semi-structured and unstructured data.

\subsubsection{Deployment Strategy}
Not many details are given on the deployment, but their proof of concept was tested on Teradata Aster DBMS where they executed their queries using SQL and SQL-MR. 

\subsubsection{Critique}
They claimed to have an end-to-end proposal for big data benchmark, but their implementation covers only a product retailer scenario which may only be relevant in that domain. Areas covered by HiBench and BigDataBench are much more diverse than theirs.

\begin{table*}[t]
  \centering
   \begin{tabular}{|c|c|>{\centering\arraybackslash} m{3cm}| >{\centering\arraybackslash} m{3cm}|c|>{\centering\arraybackslash} m{3cm}|}
 \hline
  Benchmark Suite & Workload Category & Software Stacks Used & Data Variety & Data Generator & Metrics  \\
 \hline\hline
  HiBench & Analytic & Hadoop and Spark & Text, Unstructured & Yes & Execution Time, Throughput\\ [1ex]
  \hline
  BigDataBench & Analytic and Minor NoSQL & Hadoop, Spark, MPI & Text, Graph, Unstructured, Semi Structured, Structured & Yes & Request Per Second, Operation Per Second, Operation Per Second, Execution Time \\ [1ex] 
  \hline
  BigBench & Analytic & Teradata Aster Database & Structured & Yes & Execution Time\\ [1ex] 
  \hline
  YCSB & NoSQL Databases & Cassandara, HBase, Yahoo! PNUTS, MySQL & Text, Unstructured & No & Latency in Scaling, Latency of Request \\ [1ex] 
 \hline
\end{tabular}
  \caption{Summary of Popular Cluster Benchmark Implementations}
  \label{tab:1}
\end{table*}

\subsection{Yahoo! Cloud Serving Benchmark (YCSB)}
Yahoo! Cloud Serving Benchmark is an extensible NoSQL database benchmark effort by Yahoo! with the goal of facilitating performance comparison of the new generation of NoSQL databases also known as cloud data serving systems. Four NoSQL databases were used included in this benchmark and they include Cassandra, HBase, Yahoo!'s PNUTS and a simple sharded MySQL implementation \cite{Cooper:2010:BCS:1807128.1807152}. 

\subsubsection{Software and Systems Used}
YCSB concentrates on benchmarking NoSQL databases however as of writing this paper there are various extensions of other NoSQL databases compatible with YCSB. Examples of popular database extension are MongoDB, ArrangoDB, Redis. As of writing this paper, YCSB is still in active development. The benchmark was implemented using the Java programming language.  

\subsubsection{Metrics Used}
Two major tiers were used as metrics to evaluate the performance and scalability of the systems under test. These major tiers are performance and scaling. Performance tier focuses on the latency of request when the database is under load. Latency is measured alongside throughput to access the effect of performance as the load on the server increases. The second tier of metric used is the scaling, which measures the performance as the cluster scales. Two kinds of scalability are considered. Scale up measures the performance with respect to more servers and more data. A server with a good scaleup setting should have a constant latency. Elastic speedup measures the performance of a running system while more nodes or machines are added.

\subsubsection{Workloads Used and their diversity}
No specific application domains are targeted as part of their workloads instead they examine the effect of INSERT, UPDATE, READ AND SCAN operations on the tested NoSQL databases. This approach is due to the diverse priorities for each NoSQL system.  

\subsubsection{Data Generator}
There is no specific data generator, however, the databases consist of 120 million 1Kb record, which amounts to 120 GB of data. This data is distributed across the six servers that were used in their experiment. 

\subsubsection{Deployment Strategy}
YCBS is an open-source project that can be downloaded and deployed on a cluster of NoSQL or cloud serving database. System specific configuration must be done to have it working. As of writing this paper YCSB is well extended with major implementations of NoSQL databases.

\subsubsection{Critique}
It is comparatively easier to deploy but it is dependent on Maven and Java which can be a stumbling block for nontechnical users.  

\subsubsection{Pros}
Making the benchmark extensible make it popular and there are currently YCSB is extended to cover over twenty variants of NoSQL databases.

\section{Plug And Play Bench (PAPB)}\label{stacsbench}
In order to overcome the challenges discussed in section \ref{challenges} and the critiques made in the related work section, we have implemented a proof of concept system that we coined as Plug and Play Bench (PAPB); a well integrated and easy to deploy benchmark approach, tools, and suites using containers. The implementation utilises HiBench as the benchmark suite and HDP as the cluster deployment framework. The aim is to develop, integrate and extend current benchmarking efforts into a cluster of plug and play benchmark tools. The focus is to improve the current implementation and simplify their deployment into both local and cloud clusters running big data applications. There are two main functional components of PAPB and they are: 

\begin{itemize}
    \item Plug Functional Component: Current benchmark tools and suites requires a lot of manual configuration before executing them on a big data system. Several parameters and environment variables like \texttt{HADOOP\_HOME}, \texttt{SPARK\_HOME} or home directories of the systems that you want to benchmark must be manually configured in the configuration files. Secondly, the user also has to set some optimal configuration settings for the platforms they would want to benchmark. Examples of these are \texttt{SPARK\_DRIVER} and \texttt{SPARK\_EXECUTOR\_MEMORY}. These values differ from one cluster setting to another. The goal of the Plug part is to eliminate this manual processes by auto-detecting these settings and adding them to the configuration files without much input from the user.  
    
\item Play Functional Component: The goal of the play component of PAPB is to simplify running benchmarks. Currently, for each benchmark workload, specific settings must be set in workload configuration files. Examples of these settings are data size to generate and the number of time to run a particular experiment. We simplify this process by interactively getting input from the user and impute them into the workload configuration file. 
\end{itemize}

\subsection{PAPB Architecture}\label{section:arch}
\begin{figure}[h!]
    \centering
    \includegraphics[scale=0.45]{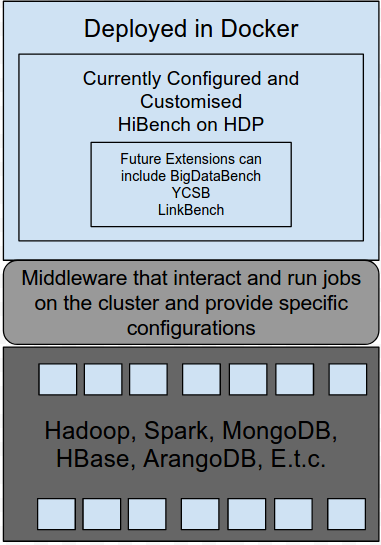}
    \caption{The propose PAPB architecture}
    \label{fig:papb}
\end{figure}

Figure \ref{fig:papb} shows the high-level architecture of PAPB. There are three main parts to this system and they are explained in the following section.

\begin{itemize}
\item Container Layer: This layer contains a mix of benchmark tools and suites. The main focus of this part is to provide configured and customised suites and tools that execute workloads on clusters without much manual configuration by the end user. In our implementation, we used the container technology and packaged HiBench based on the underlying cluster settings. PAPB was deployed on client nodes i.e. nodes within the network that can communicate with the cluster. 

\item Middleware Layer: This is the second layer and it is where the majority of this work concentrates on. It is the interface between the container layer and cluster layer. Specific settings like \texttt{HADOOP\_HOME}, \texttt{SPARK\_HOME}, name node, HDP version are obtained from the underlying cluster and these values are imputed into both the Dockerfile and configuration files of the benchmark suite. With this information, the container is built. The middleware also provides a much interactive and simplified process for running benchmarks from start to finish. 

\item Cluster Layer: From a client node, PAPB executes jobs on a cluster of machines hosting big data applications. The applications may include but are not limited to Hadoop, Spark or NoSQL databases like MongoDB, ArrangoDB, HBase etc. Using the middleware through the container, workloads are executed on the cluster and metrics sent back to the container for analysis. 
\end{itemize}

\subsection{PAPB Methodology}
In order to mitigate the challenges discussed in section \ref{challenges}, we used Docker containers to automate the deployment and integrate different benchmarking tools. This removes the majority of the manual processes involved in setting up and configuring these benchmark tools and suites. Our methodology of the simplification process follows a three-step approach as follows. First, we build the container based on the environment that the cluster runs on. This initial step is crucial as it helps in figuring out how we configure the benchmark suites that we are using. The second step involves the setup and configuration of the benchmark suites to use. The third and final step involves running the container and jobs. These processes are covered in the next section.

\subsection{Setup and Building PAPB}\label{setupandbuilding}
There are various types of cluster deployment framework and for this work, we have used Hortonworks Data Platform commonly known as HDP to deploy the cluster. Part of the setup is client nodes which are also known as edge or gateway nodes. Gateway nodes are nodes that do not store any data and they do not perform any data processing in the cluster. They are mainly used to access services, perform tasks or run jobs on the cluster. We exploit this concept by building and deploying PAPB as a running container in any of these nodes. The minimum requirement of both the normal nodes and client nodes is the ability to communicate with the cluster using password-less SSH.

\begin{algorithm}
 \caption*{Listing 1 : Building PAPB}
  \begin{algorithmic}[1]
  \label{dockerfile2}
   \State \textbf{\#First Check the cluster deployment framework and the host type}
    \State
    \State \textbf{if [ \$1=="hdp" and \$2=="normalnode"] then}
    \Indent
    \State \textbf{\#Start Building Docker file }
    \State \textbf{\#Add entries to install software prerequisite }
    \State RUN apt-get install -y [software]
    \State More dependencies...
    \State \textbf{\#Add all required ENVs to Dockerfile}
    \State \textbf{\#Example are}
     \State "HADOOP\_HOME=/usr/hdp/current/..."
     \State "PATH=\$PATH:\$HADOOP\_HOME/bin"
    \State
    \State \textbf{\#If normalnode only }
    \State rsync -avz user@sourceNode:/usr/hdp ./
    \State rsync -avz user@sourceNode:/usr/jdk64 ./
    \State rsync -avz user@sourceNode:/etc/hadoop ./
    \State 
     \State \textbf{\#Add hdfs user and login}
     \State RUN useradd hdfs
     \State RUN chown -R hdfs:hdfs HiBench
     \State USER hdfs
    \State \textbf{\#End Dockerfile building}
    \EndIndent
    \State \textbf{fi}
    \State \textbf{\#Build image from the created Dockerfile}
    \State docker build -t normalnode .
    \State \textbf{\#Start the container}
    \State docker run --name normal --network=host -p 8080:80 -it normalnode bash
  \end{algorithmic} 
\end{algorithm}

In pseudo-code Listing 1, line 3 checks the cluster deployment type and the type of node to deploy PAPB on. Line 6 installs software prerequisites. Example of such software is java, rsync, ping-utilities and Linux bc. Lines 10 to 12 set HiBench specific environment variables. It also sets the environment variables in Dockerfile needed for various client applications like Hive, Hadoop and Spark to run. Lines 15 to 17 copies the required binary and configuration files from a network node to enable the container to run as a client node. Finally, we build the container and run it.
\subsection{Deploying PAPB}
The deployment can be done on either a gateway or a normal node within the network. It is important to keep the size of the docker image as small as possible to minimise the overall deployment time. The size of the image depends on the type of node that is used to deploy PAPB. 

\subsection{Deploying PAPB on Gateway Node}
The easiest and fastest way to deploy PAPB is to deploy it on a gateway node. In our implementation, we use the power of docker volumes to map required host source directories to their corresponding guest container directories. For HiBench to work on an edge node we mapped the following source locations as volumes:\\
\begin{tcolorbox}
\begin{itemize}
\item \textbf{/usr/hdp:}  Contains all source for HDP
\item \textbf{/usr/jdk64:}  Required by HDP
\item \textbf{/HiBench:}  Contains HiBench Source directories
\item \textbf{/etc:}  Contains Hadoop, spark, hive and other configurations
\end{itemize}
\end{tcolorbox}
This helps to significantly reduce both the deployment and the size of the images. It also avoids the process of copying the files from a remote cluster node.

\subsection{Deploying PAPB Container in Normal Nodes}
In the absence of gateway nodes, software prerequisites like Java, Python, SSH must be installed on the node. Furthermore to enable the node as a client configuration files for Hadoop, Spark and other client programs must also be installed or copied from a node in the cluster into the container. This process makes the image bigger when compared to using gateway nodes.  

\subsection{Testing PAPB on Microsoft Azure}
The PAPB proof of concept was deployed and tested on Microsoft Azure cloud platform. Using HDP we deployed an eight node cluster, an edge node, and a normal node. Each node in the cluster is equipped with 112GB of memory and 16 vCPUs. We deployed and tested both Case1 and Case2 and the summaries and discussions are presented below. 
\begin{table}[t]
  \centering
   \begin{tabular}{|c|c|c|}
 \hline
  Case & Size of Image & Deployment time  \\
 \hline\hline
  Normal Node & 5.2GB & 3-4 mins \\ [1ex]
  \hline
  Edge Node & 128MB & 5 secs \\ [1ex]
  \hline
 \hline
\end{tabular}
  \caption{Summary of Deployment Approach}
  \label{tab:summary}
\end{table}

Table \ref{tab:summary} shows the size and time taken to deploy a container in each of the cases that were covered. As expected deployment in a normal node takes more time as it involves copying files from a remote server and copying them to the container. On the other hand, deploying on an edge node is straightforward because it only involves mapping relevant host directories containing binaries and configurations files to the same locations in the container.

\section{Preliminary Experimental Results}
Using the cloud environment usually incurs monetary costs since users are billed for the period that resources are being utilised. Cost metrics are not considered by existing big data benchmarking tools and therefore it would be useful to understand the monetary costs that will be incurred on running a particular workload at a particular time instance on a cluster of machines. This should not be confused with the cost analysis on cloud VM benchmarks presented in works such as \cite{iosup2011performance},\cite{zaharia2008improving},\cite{garg2013framework} and \cite{AgmonBen-Yehuda:2013:DAE:2509413.2509416} as these works focuses mostly on single cloud VM and applications are not distributed big data applications. The main goal of the experiment is to showcase both the functionality of PAPB and to demonstrate the inclusion of cost metrics in big data benchmarking. It is important to note that showcasing the functionality is crucial as it demonstrates that our idea works. In single cloud VM instance benchmarking cost metrics are used as an important measuring metrics in cloud brokerage services \cite{garg2013framework}, so it is important to replicate this in cluster benchmarking.

\begin{table}[!h]
  \centering
   \begin{tabular}{|>{\centering\arraybackslash} m{2cm}|>{\centering\arraybackslash} m{1cm}|>{\centering\arraybackslash} m{4cm}|}
 \hline
  Workload & Data Size & Description  \\
 \hline\hline
  Hadoop, Spark Aggregation & Billion Rows & Aggregates 1 Billion Rows of data in hive using Hadoop MapReduce \\ [1ex]
  \hline
 Hadoop DFSIO-Read and Write & 200GB & The workload test throughput of HDFS by doing many simultaneous read and writes.\\ [1ex]
  \hline
  Hadoop, Spark Join & 0.5 Billion, 111M Rows & The workload performs a hive natural join between two tables \\ [1ex]
  \hline
  Hadoop, Spark Kmeans & 20GB, K=10 & Performs Kmeans on 200M observations randomly generated using Uniform Distribution and Guassian Distribution. Max iteration set to 5. \\ [1ex]
  \hline
  Hadoop PageRank & 5 Million Pages, 3 Iteration & Performs the PageRank \cite{page1999pagerank} algorithm.\\ [1ex]
  \hline
  Hadoop, Spark Scan & 1 Billion Rows & Scans a 1 billion rows of record stored in Hive\\ [1ex]
  \hline
 Hadoop, Spark Word Count & 300GB & Performs the popular word count program.\\ [1ex]
  \hline
\end{tabular}
  \caption{List of Workloads Executed Using Spark and Hadoop}
  \label{tab:workloads}
\end{table}
Table \ref{tab:workloads} shows the 13 workloads that were executed on Microsoft Azure cloud platform. The workloads cover diverse sets of big data applications which include micro benchmarks, machine learning, SQL and web services. We have scaled the infrastructure from four to eight nodes to understand how performance and cost scales. To approximate the cost of running a workload we use the following relationship:

\begin{equation}
\texttt{COST} = \sum_{i=1}^{n}\frac{ \texttt{CPM * T}}{\texttt{SEC\_IN\_MNT}}
\label{eq:cost}
\end{equation}
where n = number of nodes, \texttt{CPM} is the cost (\textsterling) of a running VM per month, T is the total time taken in seconds to run a workload and \texttt{SEC\_IN\_MNT} is the total seconds in a month (approximated to 30 days). 
Each cloud service provider uses their own specific pricing model and the cost metric presented here is only relevant to the Azure platform. The Virtual Machine (VM) instances that we used is \texttt{STANDARD\_DS14\_V2} and approximately incurred a cost \textsterling821 per month when the node is running. Given the knowledge of the cost of a single VM instance, PAPB collates and calculate the cost of running that workload at the end of the entire experiment.   
We performed an experiment on an eight-node Ubuntu cluster, with one NameNode or Master and each worker node has a hard disk size of 1TB, a physical memory of 112GB and 16 CPU cores. All the nodes are deployed in the US East region data centre. Each run was executed three times and the average was taken to avoid any bias.

\begin{figure}[h!]
    \centering
    \includegraphics[scale=0.4]{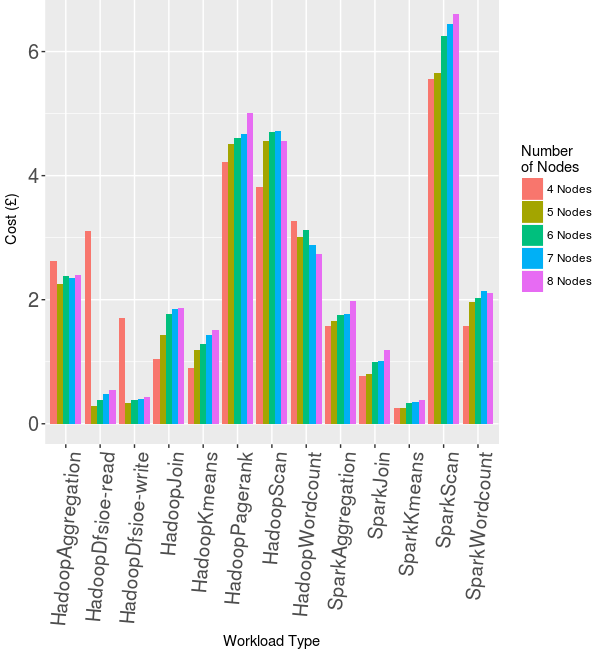}
    \caption{Cost Metrics For Different Workloads}
    \label{fig:cost}
\end{figure}
\begin{figure}[h!]
    \centering
    \includegraphics[scale=0.4]{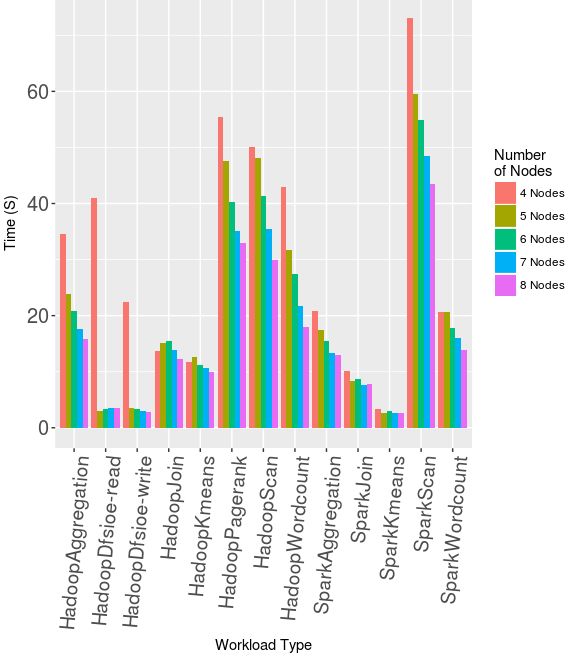}
    \caption{Execution Time Metrics For Different Workload }
    \label{fig:time}
\end{figure}

Figure \ref{fig:cost} shows the costs of running each of the benchmarks shown in Table \ref{tab:workloads} while Figure \ref{fig:time} shows the total time taken to run each of the workloads shown in Table \ref{tab:workloads} as the cluster scales from 4 nodes to 8 nodes.

From Figure \ref{fig:cost} we see that generally the cost increases as the number of nodes increases. For example, the Hadoop-Join workload has a cost of \textsterling1 and \textsterling1.8 for a four and eight nodes clusters respectively. Similarly from Figure~\ref{fig:time}, for most workloads, performance improves as the number of node increases. Example, the execution time of SparkScan workload took 75s and 45s for a four and eight nodes cluster respectively. 

\section{Conclusion and Future Work}\label{conclusion}
In this paper, we firstly compared a popular set of big data benchmarking tools by considering their strengths and weaknesses and outlined the challenges in deploying and using them. We identified that setting up and configuring current benchmarking tools is a cumbersome task and needs to be done for each individual computing environment in which they need to be used. To mitigate this challenge we proposed and implemented Plug and Play Bench, a proof of concept implementation using HDP and HiBench to simplify the setup and configuration and execution of the benchmarking process. It is important to note that the proof of concept implementation is a general idea and it was tested using HDP and HiBench but it is not tight to those two only. Other cluster deployment platform like CHD could also be used. Our implementation used the Docker container technology and dynamically build Dockerfiles based on the cluster deployment framework and underlying cloud platform. Finally, we evaluated the PAPB concept by including cost metrics to explore the monetary costs involved in executing workloads. In general, we have seen that in most workloads cost increases as the number of nodes increases but with an advantage lower execution times. The cost information can be handy for users of the PAPB, as it will help them understand the cost in relation to performance. Our future work will extend PAPB to readily support and include various cloud environments and cluster deployment frameworks. 

\balance

\ifCLASSOPTIONcompsoc
  \section*{Acknowledgments}
\else
  \section*{Acknowledgment}
\fi
This research was supported by a Microsoft Azure Award.
\label{Bibliography}

\bibliographystyle{IEEEtran}  
\bibliography{Bibliography}  

\end{document}